\documentclass[10pt, conference, compsocconf]{IEEEtran}

\usepackage[keeplastbox]{flushend} % equalizes columns on last page
\usepackage{stackengine}
\usepackage{graphicx}
\usepackage{amsmath}
\usepackage{amssymb}
\usepackage{color}
\usepackage{ifpdf}
\usepackage{blindtext}
\usepackage{float}
\usepackage[utf8]{inputenc}
\usepackage{multirow}
\usepackage{rotating}
\usepackage[caption=false]{subfig}
\usepackage{array}

\usepackage{moresize}
\usepackage{url}
\usepackage{booktabs}
\usepackage{listings}   
\usepackage{paralist}    
\usepackage{wrapfig}    
\usepackage{multirow}
\usepackage{ifpdf}
\usepackage{xspace}
\usepackage{keyval}  
\usepackage{color}
\usepackage{soul}

\usepackage{listings}
\usepackage{paralist}
\usepackage{multirow}
\usepackage[keeplastbox]{flushend} % equalizes columns on last page
\usepackage{soul}
\usepackage{amsthm}
\usepackage{thmtools}

\definecolor{listinggray}{gray}{0.95}
\definecolor{darkgray}{gray}{0.7}
\definecolor{commentgreen}{rgb}{0, 0.4, 0}
\definecolor{darkblue}{rgb}{0, 0, 0.6}
\definecolor{purple}{rgb}{0.6, 0, 0.6}
\definecolor{middleblue}{rgb}{0, 0, 0.75}
\definecolor{darkred}{rgb}{0.4, 0, 0}
\definecolor{brown}{rgb}{0.5, 0.5, 0}
\definecolor{dkgreen}{rgb}{0,0.5,0}
\definecolor{orange}{rgb}{1,.5,0}
\definecolor{dandelion}{cmyk}{0,0.29,0.84,0}

\usepackage[normalem]{ulem}
\makeatletter
\def\cyanuwave{\bgroup \markoverwith{\lower3.5\p@\hbox{\sixly \textcolor{cyan}{\char58}}}\ULon}
\def\reduwave{\bgroup \markoverwith{\lower3.5\p@\hbox{\sixly \textcolor{red}{\char58}}}\ULon}
\def\blueuwave{\bgroup \markoverwith{\lower3.5\p@\hbox{\sixly \textcolor{blue}{\char58}}}\ULon}
\font\sixly=lasy6 % does not re-load if already loaded, so no memory problem.
\makeatother

\newif\ifdraft
\drafttrue
\ifdraft
 \newcommand{\N}[1]{\textbf{NOTE: #1}\xspace}
 \newcommand{\jhanote}[1]{ {\textcolor{red} { ***SJ: #1 }}}
 \newcommand{\katznote}[1]{ {\textcolor{blue} { ***DSK: #1 }}}
 \newcommand{\amnote}[1]{ {\textcolor{dkgreen} { ***andreM: #1 }}}
 \newcommand{\mikenote}[1]{ {\textcolor{cyan} { ***Mike: #1 }}}
 \newcommand{\mtnote}[1]{ {\textcolor{dkgreen} { ***MT: #1 }}}
 \newcommand{\jonnote}[1]{ {\textcolor{dkgreen} { ***JW: #1 }}}
 \newcommand{\ynote}[1]{ {\textcolor{purple} { ***YNB: #1 }}}
 \newcommand{\note}[1]{ {\textcolor{brown} { *** #1 }}}
\else
 \newcommand{\N}[1]{}
 \newcommand{\jhanote}[1]{}
 \newcommand{\katznote}[1]{}
 \newcommand{\amnote}[1]{}
 \newcommand{\mikenote}[1]{}
 \newcommand{\mtnote}[1]{}
 \newcommand{\jonnote}[1]{}
 \newcommand{\ynote}[1]{}
 \newcommand{\note}[1]{}
\fi

\newif\ifdraft
\ifdraft
\newcommand{\terminology}[1]{ {\textcolor{red} {(Terminology used: \textbf{#1}) }}}

\newcommand{\alnote}[1]{ {\textcolor{blue} { ***andreL: #1 }}}
\newcommand{\smnote}[1]{ {\textcolor{brown} { ***sharath: #1 }}}
\newcommand{\pmnote}[1]{ {\textcolor{brown} { ***Pradeep: #1 }}}
\newcommand{\msnote}[1]{ {\textcolor{cyan} { ***mark: #1 }}}
\newcommand{\mrnote}[1]{ {\textcolor{purple} { ***melissa: #1 }}}
\else
\newcommand{\onote}[1]{}
\newcommand{\terminology}[1]{}

\newcommand{\alnote}[1]{}
\newcommand{\athotanote}[1]{}
\newcommand{\smnote}[1]{}
\newcommand{\pmnote}[1]{}
\newcommand{\msnote}[1]{}
\newcommand{\mrnote}[1]{}
\newcommand{\aznote}[1]{}
\fi

\newcommand{\B}[1]{\textbf{#1}\xspace}

\lstdefinestyle{myListing}{
  frame=single,   
  backgroundcolor=\color{listinggray},  
  language=C,       
  basicstyle=\ttfamily \footnotesize,
  breakautoindent=true,
  breaklines=true
  tabsize=2,
  captionpos=b,  
  aboveskip=0em,
  belowskip=-2em,
}      

\lstdefinestyle{myPythonListing}{
  frame=single,   
  backgroundcolor=\color{listinggray},  
  language=Python,       
  basicstyle=\ttfamily \scriptsize,
  breakautoindent=true,
  breaklines=true
  tabsize=2,
  captionpos=b,  
}

\ifpdf
\DeclareGraphicsExtensions{.pdf, .jpg, .tif}
\else
\DeclareGraphicsExtensions{.ps,  .eps, .jpg}
\fi

\setstackEOL{\#}
\setstackgap{L}{10pt}

\IEEEoverridecommandlockouts
\begin{document}

\title{Evaluating Distributed Execution of Workloads}

% \title{Analysis of Distributed Execution of Workloads}

\author{
  \IEEEauthorblockN{
    Matteo Turilli\IEEEauthorrefmark{1},
    Yadu Nand Babuji\IEEEauthorrefmark{4},
    Andre Merzky\IEEEauthorrefmark{1},
    Ming Tai Ha\IEEEauthorrefmark{1},
    Michael Wilde\IEEEauthorrefmark{4},
    Daniel S. Katz\IEEEauthorrefmark{6},
    Shantenu Jha\IEEEauthorrefmark{1}\IEEEauthorrefmark{3}
  }
  \IEEEauthorblockA{
    \IEEEauthorrefmark{1}RADICAL Laboratory, Electric and Computer Engineering, Rutgers University, New Brunswick, NJ, USA
  }
  \IEEEauthorblockA{
    \IEEEauthorrefmark{4}Computation Institute, University of Chicago \& Argonne National Laboratory, Chicago, IL, USA
  }
  \IEEEauthorblockA{
    \IEEEauthorrefmark{6}NCSA, CS, ECE, and iSchool, University of Illinois Urbana-Champaign, Urbana, IL, USA
  }
   \IEEEauthorblockA{
    \IEEEauthorrefmark{3}Computational Science Initiative, Brookhaven National Laboratory, NY, USA
    }
}

\maketitle

\begin{abstract}
Resource selection and task placement for distributed execution poses conceptual
and implementation difficulties. Although resource selection and task placement
are at the core of many tools and workflow systems, the methods are ad hoc
rather than being based on models. Consequently, partial and non-interoperable
implementations proliferate. We address both the conceptual and implementation
difficulties by experimentally characterizing diverse modalities of resource
selection and task placement. We compare the architectures and capabilities of
two systems: the AIMES middleware and Swift workflow scripting language and
runtime.  We integrate these systems to enable the distributed execution of
Swift workflows on Pilot-Jobs managed by the AIMES middleware. Our experiments
characterize and compare alternative execution strategies by measuring the time
to completion of heterogeneous uncoupled workloads executed at diverse scale and
on multiple resources. We measure the adverse effects of pilot fragmentation and
early binding of tasks to resources and the benefits of backfill scheduling
across pilots on multiple resources. We then use this insight to execute a
multi-stage workflow across five production-grade resources. We discuss the
importance and implications for other tools and workflow systems.
\end{abstract}

% ---------------------------------------------------------------------------
% INTRODUCTION
% ---------------------------------------------------------------------------
\section{Introduction}\label{sec:intro}

The distributed execution of workloads composed of many, possibly dependent
tasks poses several challenges. Effective and efficient mechanisms need to be
developed to select, acquire, and manage resources along with ways to bind,
schedule, and distribute the tasks over those resources. These mechanisms
depend on acquiring information about available resource capabilities and
workload requirements, and on selecting an appropriate set of resources on
which to execute the given workload.

Information acquisition and integration is generally feasible with
homogeneous resources and specific types of workloads, but time-dependent
availability and heterogeneous resources and workloads add significant
complexity. Users and middleware developers resort to use ``best practices,''
heuristics, or simply trial and error. This leads to suboptimal distributed
execution, due to issues such as inefficient resource and task binding,
resource under- or over-utilization, lack of data and compute co-location,
and minimal data staging preemption.

We address these limitations by devising abstractions and integrating
existing middleware. Previously, we used sequences of decisions to model the
coupling between workload requirements and resource
capabilities~\cite{turilli2016integrating}. We called these sequences
``Execution Strategies'', and used them to abstract resource selection and
how the workload's tasks are distributed on the selected resources.

We implemented execution strategies in AIMES, a pilot-based execution manager
which we used to characterize and compare alternative strategies. Differences
in the decisions composing a strategy and in the output of each decision
correspond to the selection of particular resources with specific
capabilities and to a particular binding, scheduling, and distribution of
tasks to the resources.

In this paper, we present the integration of AIMES with
Swift~\cite{wilde2011swift}. While both systems are capable of end-to-end
distributed execution of multi-task workloads, AIMES lacks some of the workflow
management capabilities offered by Swift, while Swift lacks some of AIMES's
execution coordination capabilities for diverse resources. Among the workflow
systems developed to support scientific research~\cite{taylor2007workflows}, we
choose Swift because of its modular design, its integration with at least three
pilot systems (Coasters~\cite{hategan2011coasters},
Falkon~\cite{raicu2007falkon}, and JETS~\cite{wozniak2013jets}), and access to
its developers. In principle, we could have used any other workflow system or
tool implementing distributed execution but we might have incurred in greater
engineering effort.

The integrated AIMES and Swift combine their distinctive capabilities
enabling the execution of heterogeneous workloads on heterogeneous resources.
We describe the integration of the two systems by highlighting architectural
and functional difference and similarities. We perform experiments to compare
the performance of diverse execution strategies separately with Swift and
AIMES, and we profile and emulate the execution of a real-life workflow with
the integrated systems.

Analysis of AIMES and Swift, description of their integration, and
experimental evaluations contribute towards developing a
quantitative model of distributed execution, and offer insight in how to
effectively integrate independent middleware components. Many aspects of
these advances do not depend on specific workload and resource properties.

% ---------------------------------------------------------------------------
% CONTEXT & RELATED WORK
% ---------------------------------------------------------------------------
\section{Related Work}\label{sec:rwork}

The integration of middleware components to enable large scale, distributed
computing is common. Globus~\cite{foster1997globus} and
Condor~\cite{thain2005distributed} are paradigmatic examples of systems that
aggregated several components to improve and diversify their overall
capabilities. The same applies to workflow systems, for example,
Pegasus~\cite{deelman2005pegasus}, Kepler~\cite{ludascher2006scientific},
Taverna~\cite{wolstencroft2013taverna},
JS4Cloud/DMCF~\cite{marozzo2015js4cloud}, COMPSS~\cite{lordan2014servicess} and
Swift.

The integration of independent systems not designed to be part of an existing
framework is less common for both social and technical reasons. This usually
favors the development of tailored subsystems.  Independent components have
been integrated previously; however these experiments have not been
systematically examined or characterized.

AIMES uses a pilot system called RADICAL-Pilot (RP) to execute
workloads~\cite{merzky2015radical}. A large number of pilot systems have been
developed for specific resources or type of
workloads~\cite{turilli2017comprehensive}, but RADICAL-Pilot has been
designed to be as generic and open as feasible.

Some of the issues underlying the characterization and comparison of
different ways to distribute workload execution across heterogeneous
resources have been studied in
detail~\cite{blythe2003role,adler1995distributed}. The same has been done
with dynamic scheduling of
workflows~\cite{prodan2005dynamic,fard2013truthful}. Nonetheless, a
comprehensive characterization of the performance tradeoffs among alternative
models of distribution is still unavailable, especially when related to
pilot-enabled systems. This paper contributes towards the development of this
research topic.

% ---------------------------------------------------------------------------
% ARCHITECTURES
% ---------------------------------------------------------------------------
\section{Architectures}\label{sec:archs}

An execution strategy is the sequence of decisions that have to be made to
execute a workload on resources. Each decision selects among alternative
choices: actions, entities, or attributes, depending on the selection
process, workload, and resources. A sequence of choices is a realization of
an execution strategy.

Each decision of an execution strategy is based on evaluating how alternative
choices satisfy one or more metrics, which in turn are based on models of how
choices relate to metrics. These metrics may relate to properties of the
workload (e.g., duration, size, degree of concurrency), or properties of
resources (e.g., hardware type, location), or they might be federation and
economic models (e.g., Grid, Clouds, energy consumption, cost of resources,
allocation).

For example, for the `time-to-completion ($TTC$) of a workload' metric,
deciding how many and which resources to use relies on models of each
resource's compute performance. Decisions are made by users (e.g., via
configuration files), programmers (e.g., via how the application is written),
or algorithms (e.g., via runtime decisions).

We compare AIMES and Swift qualitatively by looking at the decisions of their
execution strategies to see differences in capabilities and their
implementations. Quantitatively, we compare execution strategies against how
closely the \(TTC\) of executed workloads approximates their ideal \(TTC\).

% ---------------------------------------------------------------------------
\subsection{AIMES}\label{sub:archs-aimes}

\begin{figure}[t]
  \subfloat[\label{subfig:aimes-arch}]{%
    \includegraphics[width=0.25\textwidth]{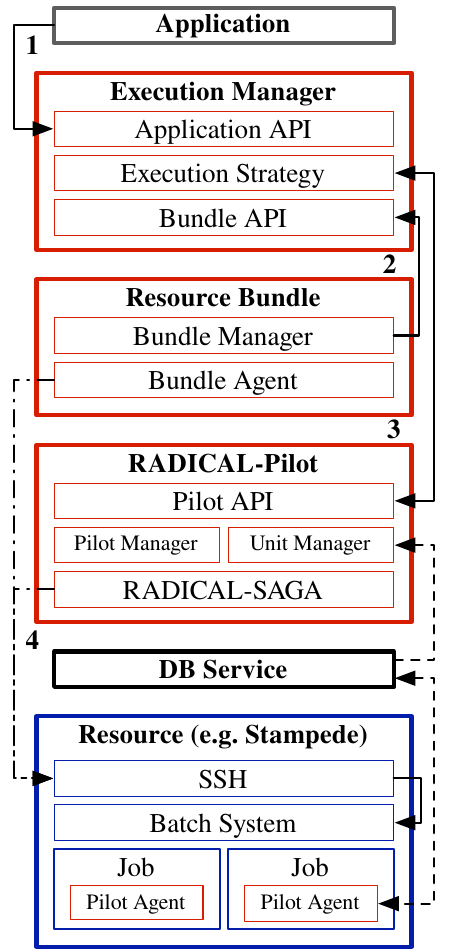}
  }
  \subfloat[\label{subfig:swift-arch}]{%
    \includegraphics[width=0.25\textwidth]{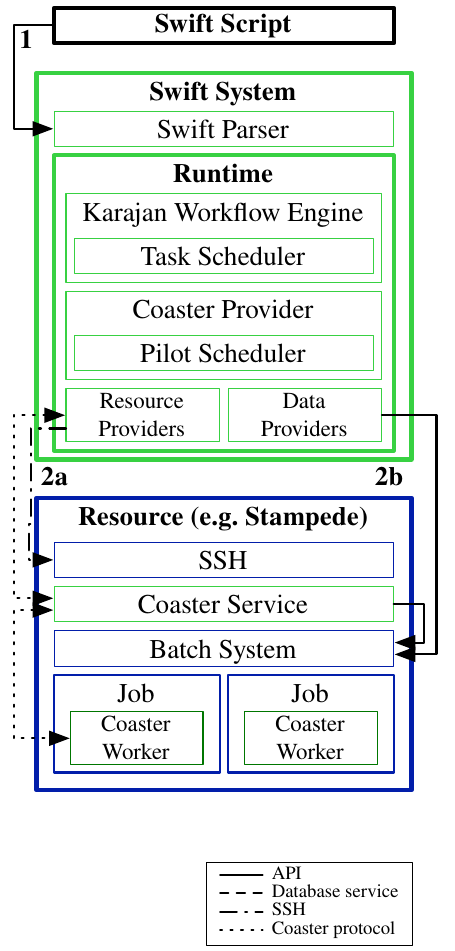}
  }
  \caption{ \B{(a)} AIMES and \B{(b)} Swift detailed architecture diagrams.
            Red boxes indicate AIMES components, green Swift components, blue
            resources, and black third-party components.\label{fig:archs}}
\end{figure}

AIMES enables the execution of distributed applications on HPC and HTC systems.
The architecture of AIMES has three main components: an Execution Manager, a
pilot system (`RADICAL-Pilot'); and a resource information system named `Bundle'
(Fig.~\ref{subfig:aimes-arch}). The Execution Manager collects information about
workload requirements from the application layer (Fig.~\ref{subfig:aimes-arch},
1) and both static and dynamic information about resource capabilities from
Bundle (Fig.~\ref{subfig:aimes-arch}, 2).

The Execution Manager integrates information and realizes an execution strategy
based on user configuration and dedicated algorithms. Currently, users define
the percentage of maximum concurrency with which to execute the workload's tasks
and the percentage of available resources that should be used. Algorithms
calculate the amount of resources and the time for which they have to be
available, how to partition the resources across pilots, and how to distribute
pilots across resources.

This decision process has been empirically tailored to minimize the \(TTC\) of
bag of tasks (BoT) and multi-stage BoT where subsets of the BoT's tasks have to
be sequentially executed due to data dependences among tasks. The Execution
Manager is designed to support the development of alternative decision
strategies and optimization metrics. Thanks to a modular design, every decision
of an execution strategy can be implemented independently and multiple decisions
can be statically (and in the future dynamically) grouped into sequences on the
basis of the decisions' interdependences and priority.

The Execution Manager enacts execution strategies via RADICAL-Pilot
(Fig.~\ref{subfig:aimes-arch}, 3). A Pilot Manager describes, instantiates, and
monitors pilots. Each pilot is submitted to the resource as a job and, once
scheduled, executes a Pilot Agent. A Unit Manager translates application tasks
into Compute Units, which it then schedules and executes on a Pilot Agent.
RADICAL-Pilot uses an interoperability layer called RADICAL-SAGA to access the
resources' batch systems (Fig.~\ref{subfig:aimes-arch}, 4). RADICAL-SAGA enables
pilot submission and data staging over multiple interfaces.

The use of a pilot system enables late binding of workloads to pilots. Late
binding is the ability to utilize pilots dynamically, i.e., the workload is
distributed onto pilots only when the pilots are effectively available.
RADICAL-Pilot is capable of distributing the workload across pilots instantiated
on diverse resources. This enables late binding to both pilots and resources: a
workload is submitted to a specific resource only when a pilot on that resource
is available.

AIMES is implemented as four Python modules: aimes.emgr, the Execution Manager;
aimes.bundle, the Bundle information system; radical.pilot, the RADICAL-Pilot
pilot system; and radical.saga, the RADICAL-SAGA interoperability layer. Each
module exposes well-defined or standardized interfaces and can thus be used
independently. For example, both RADICAL-SAGA and RADICAL-Pilot are
independently used by diverse scientific communities.

The AIMES components communicate globally via a database service, and locally
via distributed messaging. The AIMES Execution Manager and RADICAL-Pilot act in
a Master/Worker pattern, as do the RADICAL-Pilot's managers with regard to the
resources' batch system and the Pilot Agents. This decouples the global and
local states of resource selection and execution management. AIMES uses late
binding to both pilots and resources to select resources independent of how they
will be used for execution.

AIMES maintains a global view of workload execution. The requirements of the
workload's tasks are evaluated before any pilot is described and assigned to a
resource. Once evaluated, tasks are scheduled into a global queue controlled by
the RADICAL-Pilot Unit Manager. Thus, AIMES can operate on multiple pilots on
multiple resources with global scheduling algorithms. These algorithms can, for
example, maximize overall resource utilization, prioritize resources, or
evaluate resource affinities.

% ---------------------------------------------------------------------------
\subsection{Swift}\label{subsec:archs-swift}

Swift is both a parallel scripting language and a runtime system, used to
compose and orchestrate software applications or high-level library functions.
There are two implementations of Swift: the original `Swift/K'
system~\cite{wilde2011swift}, primarily intended for local and distributed
execution of workflows composed of file-passing applications, and a newer
Swift/T system designed to additionally support object-passing functions by
running tasks in-memory on large-scale parallel systems. Here we focus on
Swift/K and refer to that system as `Swift,' but a similar analysis,
integration, and set of experiments could be performed with Swift/T.

The Swift architecture has two main components: an interpreter of the Swift
language and a runtime system that executes Swift programs on parallel or
distributed resources (Fig.~\ref{subfig:swift-arch}). Swift programs are
executed in a two-stage manner. First, a Swift script is parsed and converted
into Karajan~\cite{von2007java} (Fig.~\ref{subfig:swift-arch}, 1), a
declarative-style language with strict evaluation. The Karajan script is then
executed by the Karajan workflow engine, which in turn may invoke Swift-specific
primitive functions. The Swift runtime system uses a plug-in framework with
\textit{providers} that submit app tasks to diverse computing resources
(Fig.~\ref{subfig:swift-arch}, 2a and 2b) and that move files to and from those
tasks.

The `Coaster' system~\cite{hategan2011coasters} is a provider that employs the
pilot abstraction~\cite{turilli2017comprehensive} to allocate computing
resources. Pilot are submitted to a computing site's resource manager (RM) as
jobs and, once instantiated, compute node agents (Coaster workers) launch tasks
on the nodes and transfer files to and from the node's local filesystem. In this
paper we only consider Swift program execution using the Coaster pilot provider;
this is the most common way in which Swift is currently used and is analogous to
how AIMES uses RADICAL-Pilots.

Swift schedules app tasks in a multi-stage manner, with responsibility
partitioned between the Karajan runtime system and the Coaster pilot
execution provider. In terms of execution strategies, this process can be
thought of as a sequence of decisions, starting with higher-level scheduling
decisions made during language interpretation and ending with lower level
decisions made in the pilot scheduler (Fig.~\ref{subfig:swift-arch}).

Swift attempts to schedule tasks to a computing site, controlled by two
levels of per-site throttles. At the higher level, each site can, at any
given time, accept a certain number of app tasks. At a lower level, Swift
also effectively throttles the rate at which tasks can be submitted to the
site by limiting the number of concurrently active submission and file
management tasks on a per-site basis. Once an app task clears all applicable
throttles, the task is queued to the site.

The total number of tasks that can be concurrently queued to a given site can
be fixed by the user, or can be dynamically controlled by an automatic site
selection algorithm that considers three factors: the number of tasks already
queued at the site, the rate at which the site is successfully completing
jobs, and the rate (ideally zero) at which jobs may be failing at the site.
The dynamic site selection algorithm seeks to simultaneously balance work
between sites, assign work based on site productivity, and withhold work to
reduce the chances of an app task failing at a site.

Once an app task is assigned to a site, it is enqueued to Swift's Coaster
(pilot) provider and it enters a per-site queue. The pilot scheduler
periodically visits each site's queue to determine the compute node resources
needed for that site. The parameters that govern this resource allocation are
specified by the user as the maximum number of nodes that can be allocated
for the site, and the manner in which that allocation of nodes is grouped
into pilot jobs that are enqueued to the site's resource manager.

Within these constraints, at each scheduling interval the pilot scheduler
performs a box-sizing-and-packing algorithm to determine what size pilot jobs
need to be submitted (if any) and then what jobs should be packed into each
box~\cite{hategan2011coasters}. Once pilot jobs are started by a RM and
compute nodes are provided to the pilot scheduler, the pilot scheduler places
app tasks into task slots on the compute nodes through the node worker agents
started by the pilot jobs.

% ---------------------------------------------------------------------------
% COMPARISON
% ---------------------------------------------------------------------------
\section{Comparison}\label{sec:comparison}

AIMES and Swift have overlapping scopes. As end-to-end systems, they both enable
users to specify multi-task workloads and execute them on diverse types of
resources: Grid, HPC, Cloud. Functionally, users specify and execute
applications in different ways on the two systems. These differences are mainly
related to workload specification, resource selection, resource partitioning
into pilots, pilot and task binding, and task execution. Here we summarize these
differences, highlighting those most relevant to the integration of the two
systems, which we will describe in the next section.

\paragraph{Workload specification} AIMES does not offer a native workload
description language; users have to use third party languages. These are
supported via interface modules that take a workload description as input and
return the description of a set of tasks as output. Internally, AIMES represents
tasks as compute units (CU), the same data structure used by RADICAL-Pilot. CUs
have a set of predefined properties including, for example, executable,
executable's arguments, number of cores, or message passing interface. Input and
output files can be specified for each CU, but limited support is given to the
specification of inter-task dependences and to the grouping of tasks in stages
or bulks.

Swift is specifically aimed at workflows. The Swift programming language is
designed to specify tasks as functions or external executables, alongside their
data dependences. The Swift language is implicitly parallel and it includes
control structures to describe, for example, the mapping of variables to
physical files, the execution of groups of tasks, or the remote execution of
specific functions.

\paragraph{Resource selection} AIMES uses Bundles to obtain static and dynamic
information about the capabilities and availability of target resources. This
information is polled from the Bundle database via a dedicated API, enabling
AIMES to select resources on the base of both historical and real-time
evaluations. For example, a resource may be chosen because its compute and data
capabilities satisfy the requirements of the given workload but also because
that specific resource has been historically reliable.

Swift enables users to select resources via a configuration file. It can contain
an entry for each resource, specifying the parameters required by the execution
of their workflow. Users can set the type of Swift provider they want to use
with the resource, the address of the resource's endpoint, the type of resource
manager, the data staging modalities, a working directory, and also parameters
that determine the parallelism and the concurrency of tasks execution.

\paragraph{Resource partitioning} Both AIMES and Swift can submit single or
multiple pilots with variable cores and walltime to one or more resource. The
two systems implement this differently: AIMES derives the number, binding, size,
and duration of the pilots based on the given execution strategy; Swift uses
user-provided configuration files or the resource partitioning algorithm
described in~\S\ref{subsec:archs-swift}.

\paragraph{Resource and task binding} AIMES enables early binding of pilots
to sites and late binding of tasks to pilots. Pilots abstract the
capabilities of the resource on which they are instantiated and tasks are
bound to a pilot depending on whether the pilot's capabilities satisfy the
tasks' requirements. For example, a task requiring 128 cores is bound to a
pilot with at least that number of free cores, but the same pilot might not
be used for tasks requiring large memory.

AIMES binds tasks only to pilots, not to resources. Tasks are bound to the
first available and suitable pilot, independent from the resource on which
the pilot has been instantiated. As such, in multi-site executions, not all
queued pilots need to be used to execute a workload. Given enough time
difference between the first pilot to become active and all the others, all
the tasks of a workload could be executed on the first pilot that becomes
active.

Late binding to pilots introduces both positive and negative performance
trade-offs. For example, distributing multiple pilots across multiple
resources reduces the overall time spent in a queue waiting for a pilot to
become active~\cite{turilli2016integrating}. Conversely, late binding to
pilots requires more time be spent staging and replicating data across
multiple resources.

Swift currently binds some tasks to resources when a site selection algorithm
is used, and binds all tasks to resources when the binding is user
configured. For multi-site execution, at least a few tasks need to be
executed on all the given resources. This behavior could be changed by
implementing a different scheduling algorithm for the Pilot Scheduler of the
Coaster Provider (Fig.~\ref{subfig:swift-arch}).

\paragraph{Task execution} AIMES executes workloads by enacting one of the
Execution Manager's execution strategies. The currently available execution
strategy requires that all the tasks of the workload are known before
starting the workload execution. Consequently, AIMES cannot execute workloads
in which tasks become available after execution has already started.
Currently, AIMES does not support task and data replication, or pilot fault
tolerance.

The isolation between the Swift interpreter, workflow engine, and resource
providers guarantees the separation of concern between task specification and
execution. Swift also enables the separation between the provisioning of
tasks and instantiation of pilots. Pilots can be reused when available and
scheduled while the the execution is in progress depending, for example, on
how a specific site is performing.

% ---------------------------------------------------------------------------
% INTEGRATION
% ---------------------------------------------------------------------------
\section{Integration}\label{sec:integration}

We integrated AIMES and Swift to combine their distinguishing
functionalities. This is technically challenging mostly due to the different
programming languages used for the two systems. Functionally, the main issues
are to account for the differences in how workloads are described and
executed, and in the capabilities of the two systems. Specifically, the
handling of exceptions and failure, the logging mechanisms, and the state
transitions have to be reconciled.

The goal of this integration is to compare of the execution strategies
implemented by Swift, AIMES, and their combination, not to explore diverse
integrative architectures nor to choose the best architecture for a specific
metric. For this reason, we developed a prototype of the two-system
integration, focusing on those capabilities specifically required by workload
execution. Robustness, fault-tolerance, and flexibility are not primary
concerns of our prototype.

A main common point of AIMES and Swift is a task-oriented application model
(\S\ref{sec:comparison}). Both systems assume that a set of tasks is
described and then scheduled for execution on suitable resources. The two
systems use different task descriptions but both specify the executable to be
run on the chosen resource, its arguments, and its inputs and outputs. AIMES
also requires specifying the number of cores required by each task and an
estimate of the time the task will take to execute. We added this information
to the Swift task description.

\begin{figure}[t]
    \centering
    \includegraphics[width=0.4\textwidth]{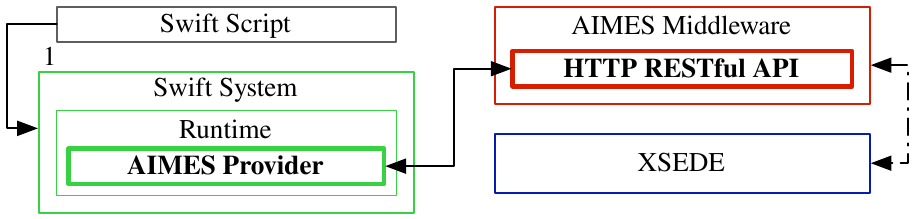}
    \caption{AIMES and Swift integrated architecture. A RESTful API and
             dedicated connectors have been developed to support
             communication and coordination among the two systems. Only the
             components developed for the integration are displayed. The
             remaining components, colors, and arrows are as in
             Fig.~\ref{fig:archs}.\label{fig:aimes-swift-arch}}
\end{figure}

We developed an interface to enable the two systems to exchange task
descriptions (Fig.~\ref{fig:aimes-swift-arch}). The interface takes Swift
task descriptions as input, translates them to AIMES task descriptions, and
outputs them to the AIMES Execution Manager. The interface was implemented
via a dedicated provider for Swift (Fig.~\ref{fig:aimes-swift-arch}, AIMES
provider) and a HTTP-based RESTful API\@. The use of REST helps encapsulate
the resource provisioning logic by implementing it as a persistent service.
This approach also helps to bridge differences between Java and Python-based
services. The input and output data of the RESTful API are formatted in JSON,
describing the application tasks entirely based on their interfaces, which
thus eliminates any dependency on the type of language used to develop each
system.

AIMES and Swift have different execution models. The Swift provider has no
information about the global state of the workflow, with the total number of
tasks unknown, as is whether tasks are or will be grouped in stages, with or
without current or future data dependences. AIMES requires the whole workload
to be known before starting its execution. The RESTful interface enables
Swift's AIMES provider to submit tasks for execution as soon as they are
provided by the Karajan Workflow Engine (Fig.~\ref{fig:aimes-swift-arch}, 1).
Meanwhile, AIMES can wait to execute tasks until the task submission rate
falls below a certain rate (Fig.~\ref{fig:aimes-swift-arch}, 2).

This allows AIMES to effectively execute portions of workflows as if they
were independent workloads, i.e., a group of independent tasks
(Fig.~\ref{fig:aimes-swift-arch}, 3). For example, given a workflow with two
stages, where the input of the second stage tasks are the output of the first
stage tasks, Swift submits all the first stage tasks to the RESTful interface
and then waits for their output to be available. AIMES does not need
information about whether the given group of tasks belongs to a stage, or
whether a second stage exists. AIMES monitors the task submission rate to the
RESTful interface and, once a configurable amount of time has passed since
the last task submission, it executes all the submitted tasks as a
self-contained workload.

% ---------------------------------------------------------------------------
% EXPERIMENTS
% ---------------------------------------------------------------------------
\section{Experiments}\label{sec:exp}

We designed two sets of experiments to characterize and compare execution
strategies. We executed BoTs and workflows on XSEDE and NCSA resources,
studying the effects of strategies' decisions on \(TTC\). The first set of
experiments was executed with Swift and AIMES separately, and the second with
Swift and AIMES integrated. All experimental data, code, and analysis are
publicly available~\cite{aimes_swift_experiments_url}.

Our experiments serve four purposes: (i) to investigate alternative execution
strategies to execute workloads of different sizes and types across multiple
resources; (ii) to compare the tradeoffs imposed by these execution
strategies on \(TTC\); (iii) to outline how design features and configuration
parameters enable execution strategies; and (iv) to illustrate how the
integration of AIMES and Swift supports profiling and emulation of real-life
workflows on distributed and heterogeneous resources.

Each experiment executes an increasingly large BoT or workflow on two to four
XSEDE resources and NCSA's Blue Waters, depending on AIMES, Swift, or their
integrated capabilities. We use task and pilot concurrency within and across
resources, measuring realistic overheads and tradeoffs on production
resources. In this way, our experiments are representative of the conditions
under which users execute scientific workloads and workflows.

We designed state models for the AIMES, Swift, and integrated middleware,
defining the \(TTC\) of our experiments as: \(TTC = T_x + T_w\). According to
these state models, \(T_x\) is computed as the sum of the times required by
task scheduling, bootstrapping, staging input files (when needed), execution,
staging output files (when needed), and shutdown as performed by the pilot on
which the task is executed. \(T_w\) is computed as the sum of the times
required by the AIMES, Swift, or integrated middleware, and queuing pilots on
the target resources.

We developed data analysis toolkits to timestamp the start and end of each
state of the AIMES, Swift, and integrated
middleware~\cite{aimes_swift_experiments_url}. Parsers, filters, and
aggregators were used to measure the duration of all the states contributing
to \(T_x\) and \(T_w\) and, therefore, to \(TTC\).

We define the performance of an execution strategy as: \(P_{ES} =
(\frac{TTC_i}{TTC}) \times 100\). \(TTC_i\) is the ideal \(TTC\), calculated
by assuming maximal task concurrency for the given experimental conditions.
Task concurrency depends on the number, size, and duration of the pilots on
which the tasks are executed, while task duration is known by design,
profiling, or observation. In practice, \(TTC_i\) can never be achieved as
middleware and resources always impose some overhead, however minimal.

% ---------------------------------------------------------------------------
\subsection{Standalone: AIMES and Swift}\label{ssec:exp_aimes_swift}

We performed experiments with Swift and AIMES separately, adopting four
execution strategies and executing BoTs of varying size
(Table~\ref{table:exp_swift}). The first and second experiment were performed
with Swift, the third and fourth with AIMES\@. Swift's experiments were
performed on Stampede and Gordon, the two XSEDE resources then supported by
that system. AIMES's experiments were performed on Stampede and Gordon, and
on Stampede, Gordon, SuperMIC, and Comet.

\begin{table*}
  \centering
  \caption{ Experiments performed with Swift (1 and 2) and AIMES (3 and
    4). T = Task; R = Resource; P = Pilot.}\label{table:exp_swift}
  \begin{tabular}{clllllllll}
    \toprule
      \multicolumn{1}{c}{
        \multirow{2}{*}[-0.4em]{\bfseries \Longstack{Experiment\# ID}}} &
      \multicolumn{1}{c}{
        \multirow{2}{*}[-0.4em]{\bfseries \Longstack{System\# Name}}}   &
      \multicolumn{2}{c}{\B{Workload}}                                  &
      \multicolumn{6}{c}{\B{Execution Strategy}} \\
    \cmidrule(r){3-4}
    \cmidrule(r){5-10}
      &  & \multicolumn{1}{c}{\B{\#T}} & \B{T Duration} & \B{\#R} & \B{R binding} & \B{P binding} & \B{P walltime} & \B{P cores} & \B{\#P} \\
    \midrule
      \B{1}                    &  % Experiment ID
        \multirow{2}{*}{Swift} &  % System Name
        $8, 32, 256, 2,048$        &  % Number of Tasks
        $20$~m                  &  % Task Duration
        $2$                    &  % Execution Strategy: # Resources
        early                  &  % Execution Strategy: Resource binding
        late                   &  % Execution Strategy: Pilot binding
        $25$~m                  &  % Execution Strategy: Pilot Walltime
        $16$                   &  % Execution Strategy: Pilot Cores
        $40$                   \\ % Execution Strategy: #Pilots
      \B{2}                    &  % Experiment ID
                               &  % System Name
        $32, 128, 512, 1,024, 2,048$ &  % Number of Tasks
        $20$~m                  &  % Task Duration
        $2$                    &  % Execution Strategy: # Resources
        early                  &  % Execution Strategy: Resource binding
        late                   &  % Execution Strategy: Pilot binding
        $75$--$255$~m           &  % Execution Strategy: Pilot Walltime
        $16$                   &  % Execution Strategy: Pilot Cores
        $2$--$32$               \\ % Execution Strategy: #Pilots
     \midrule
      \B{3}                    &  % Experiment ID
        \multirow{2}{*}{AIMES} &  % System Name
        $8, 32, 256, 2,048$        &  % Number of Tasks
        $20$~m                  &  % Task Duration
        $2$                    &  % Execution Strategy: # Resources
        late                   &  % Execution Strategy: Resource binding
        late                   &  % Execution Strategy: Pilot binding
        $40$~m                   &  % Execution Strategy: Pilot Walltime
        $4$--$1024$             &  % Execution Strategy: Pilot Cores
        $2$                    \\ % Execution Strategy: #Pilots
      \B{4}                    &  % Experiment ID
                               &  % System Name
        $8, 32, 256, 2,048$        &  % Number of Tasks
        $20$~m                  &  % Task Duration
        $4$                    &  % Execution Strategy: # Resources
        late                   &  % Execution Strategy: Resource binding
        late                   &  % Execution Strategy: Pilot binding
        $80$~m                   &  % Execution Strategy: Pilot Walltime
        $2$--$512$              &  % Execution Strategy: Pilot Cores
        $4$                    \\ % Execution Strategy: #Pilots
    \bottomrule
  \end{tabular}
\end{table*}

The execution strategy of Experiment 1 (Table~\ref{table:exp_swift})
maximizes the number of pilots executing the BoT and minimizes the walltime
requested for each pilot.  Up to 20 pilots (the maximum number of
concurrently submitted pilots reliably supported by Swift on Stampede and
Gordon), each with 16 cores, were queued on Stampede and on Gordon, enabling
a theoretical total of 640 concurrent task executions. Each pilot had enough
walltime to execute up to 16 20-minutes long tasks. Each pilot was canceled
after being active for 25 minutes, accounting for bootstrap and shutdown
overheads.

The effects of this strategy on the \(TTC\) of the BoT depends on the
availability of resources and the overheads introduced by managing up to 40
concurrent pilots, 20 for each resource.  Resource availability is determined
by how long the pilots are queued in the site's RM that, in turn, depends on
load and fair share policies. Policies favoring short walltime may result in
rapid pilot turnover, near to maximal concurrency, and therefore near to
ideal \(TTC\). On the contrary, large pilot turnover may also increase
management overheads, preventing full use of the available pilots and, as a
consequence, increasing the \(TTC\) of the workload.

\begin{figure}
  \centering
  \includegraphics[width=\linewidth]{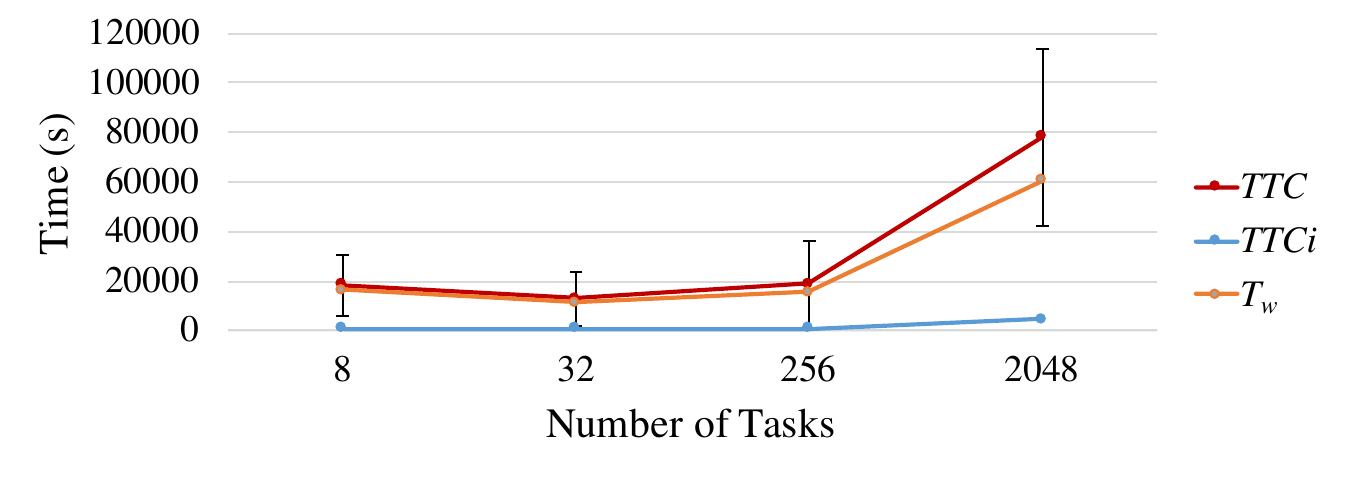}
  \caption{ \B{Experiment 1}. Average \(TTC\) and \(T_w\) for bags of 8, 32,
            256, 2,048 tasks executed on a total of between 40 and 160 pilots
            instantiated in groups of 20 pilots on both Stampede and Gordon.
            \(TTC_i\) of 8, 32, and 256 tasks is 1,200~s, of 2,048 is
            4,800~s.
            \(T_w\) is the major time contribution to \(TTC\); the error bars
            of \(T_w\) (not shown for clarity) are of the same order as those
            of TTC\@. The large relative error of \(TTC\) depends upon the
            large fluctuation of queue time for each pilot on both
            resources.}\label{fig:exp_swift_1}
\end{figure}

The blue line in Fig.~\ref{fig:exp_swift_1} shows \(TTC_i\) (ideal \(TTC\))
for bags of 8, 32, 256, and 2048 tasks. The \(TTC_i\) of 8, 32, and 256 tasks
is 20 minutes: all the task could be executed concurrently by less than 40
16-cores pilots and the duration of each task is 20 minutes. The \(TTC_i\) of
2048 tasks is instead 80 minutes: the given set of pilots can execute 640
tasks every 20 minutes.

The first row of Table~\ref{table:es_efficiency} shows the average
performance \(P_{ES}\) of the execution strategy of Experiment 1. Performance
appears to be inversely proportional to the size of the BoT, likely because
the execution of smaller BoT requires fewer active pilots on each resource.
Fig.~\ref{fig:exp_swift_1} confirms this, illustrating how most of \(TTC\) is
spent in \(T_w\). Our analysis of the components of \(T_w\) shows that the
time spent waiting for the pilots to become active on both Stampede and
Gordon queues was the dominant element of \(T_w\).

\begin{table}
  \centering
  \caption{ Average performance (\(P_{ES}\)) of the execution strategies for
    the \(TTC\) metric.}\label{table:es_efficiency}
  \begin{tabular}{clllllll}
    \toprule
      \multicolumn{1}{c}{
        \multirow{2}{*}[-0.4em]{\bfseries \Longstack{Experiment\# ID}}} & % 1a
      \multicolumn{7}{c}{\B{\(P_{ES}\) per number of Tasks}} \\
    \cmidrule(r){2-8}
                      &  % 1b
      \B{8}           &  % 2
      \B{32}          &  % 4
      \B{128}         &  % 6
      \B{256}         &  % 8
      \B{512}         &  % 10
      \B{1024}        &  % 12
      \B{2048}        \\  % 14
    \midrule
      \B{1}           &  % ID           1b
        $22\%$        &  % 8    %       2
        $31\%$        &  % 32   %       4
        $-$           &  % 128  %       6
        $19\%$        &  % 256  %       8
        $-$           &  % 512  %       10
        $-$           &  % 1024 %       12
        $7\%$         \\ % 2048 %       14
      \B{2}           &  % ID           1b
        $-$           &  % 8    %       2
        $7\%$         &  % 32   %       4
        $4\%$         &  % 128  %       6
        $-$           &  % 256  %       8
        $3\%$         &  % 512  %       10
        $4\%$         &  % 1024 %       12
        $11\%$        \\ % 2048 %       14
      \B{3}           &  % ID           1b
        $62\%$        &  % 8    %       2
        $77\%$        &  % 32   %       4
        $-$           &  % 128  %       6
        $61\%$        &  % 256  %       8
        $-$           &  % 512  %       10
        $-$           &  % 1024 %       12
        $31\%$        \\ % 2048 %       14
      \B{4}           &  % ID           1b
        $46\%$        &  % 8    %       2
        $47\%$        &  % 32   %       4
        $-$           &  % 128  %       6
        $46\%$        &  % 256  %       8
        $-$           &  % 512  %       10
        $-$           &  % 1024 %       12
        $33\%$        \\ % 2048 %       14
      \B{Integrated}  &  % ID           1b
        $-$           &  % 8    %       2
        $-$           &  % 32   %       4
        $-$           &  % 128  %       6
        $17\%$        &  % 256  %       8
        $12\%$        &  % 512  %       10
        $8\%$         &  % 1024 %       12
        $4\%$         \\ % 2048 %       14
    \bottomrule
  \end{tabular}
\end{table}

When compared to Experiment 1, the execution strategy in Experiment 2 reduces
the number of pilots to between 2 and 32 and increases their walltime to
between 75 and 225 minutes (Table~\ref{table:exp_swift}). The smallest BoT
increases to 32 as 8 tasks could be executed on a pilot on a single resource
while 2048 tasks is kept as the upper boundary. Though the BoT sizes for
Experiment 1 and 2 are not the same, they are interleaved, which permits
linear extrapolation and thus the claim that the \(TTC\) values overlap.

The strategy of Experiment 2 is an attempt to reduce the \(T_w\) observed in
Experiment 1 by limiting the number of pilots used to execute the BoT. This
is done by: (i) reducing the number of concurrent pilots queued on each
resource from 20 to 16 for a maximum of 256 concurrent task execution; (ii)
Queuing less than 16 pilots for a resource when less than 256 concurrent
cores are required to execute the BoT\@; and (iii) increasing the maximum
duration of the pilots so as to allow pilot reuse when supported by the Swift
task scheduler.

\begin{figure}
  \centering
  \includegraphics[width=\linewidth]{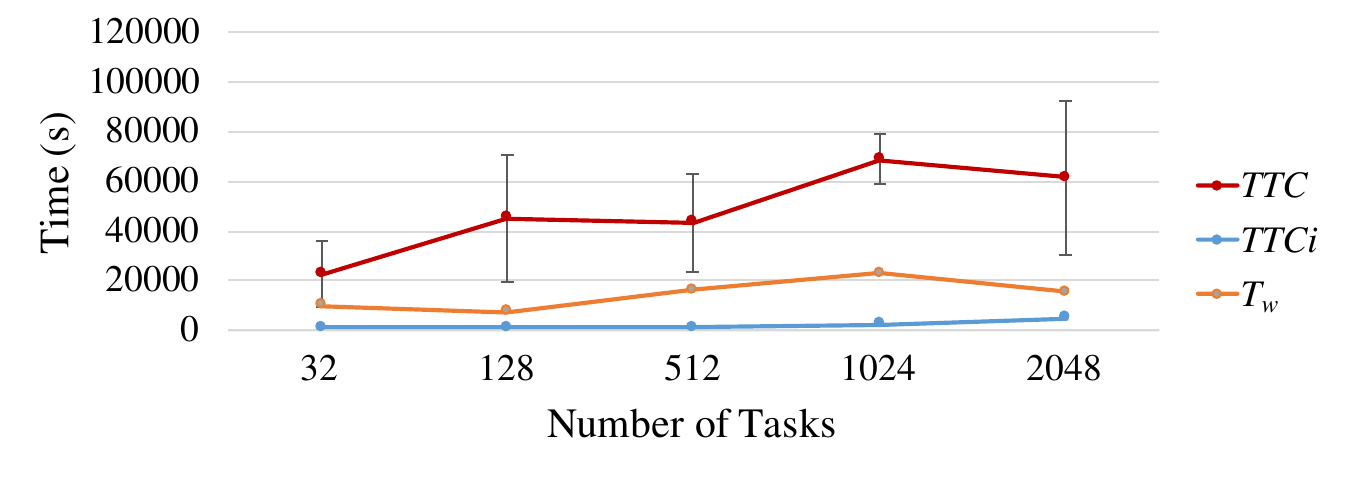}
  \caption{ \B{Experiment 2}. Average \(TTC\) and \(T_w\) for bags of 32, 128,
            512, 1,024, 2,048 tasks executed on a total of between 2 and 36
            pilots instantiated on both Stampede and Gordon. \(TTC_i\) of 32,
            128, and 512 tasks is 1200~s, of 1,024 is 2400~s, of 2,048 is 4800~s.
            As in Fig.~\ref{fig:exp_swift_1}, the error bars of \(T_w\) (not
            shown for clarity) are of the same order as those of TTC, the
            large values of which depend upon of fluctuations of the queue
            time (\(T_w\)) for each pilot on both
            resources.}\label{fig:exp_swift_2}
\end{figure}

The second row of Table~\ref{table:es_efficiency} shows that the average
\(P_{ES}\) of Experiment 2 improves for larger BoTs but worsens for smaller
BoTs, relative to Experiment 1.  It is difficult to discern this from the
\(TTC\) averages and the error bars in Fig.~\ref{fig:exp_swift_2}
and~\ref{fig:exp_swift_1}, as the two strategies are essentially equivalent.
However, with more repetitions, a reduction in errors may indeed indicate
that the execution strategy of Experiment 2 performs better.

Analogous to Experiment 1, \(T_w\) is a significant component of \(TTC\) of
Experiment 2 (Fig.~\ref{fig:exp_swift_2}),  which is a consequence of both
experiments being dependent on the pilot with the longest queue time to
become active. Unlike Experiment 1 however, there are relevant systematic
errors in \(T_w\) for Experiment 2.

Configuration parameters used to determine the execution strategies in Swift do
not fully control how the tasks are distributed across resources, since the
Swift task scheduler (Fig.~\ref{subfig:swift-arch}) operates this distribution,
always binding at least a few tasks to each resource. Experiments 1 and 2 show
that this behavior affects \(TTC\) on resources with a variable queue time. This
behavior could be changed by implementing a different task scheduler, based on
matching task requirements to pilot capabilities, as has been implemented in
AIMES\@. We designed Experiment 3 and 4 to characterize and measure execution
strategies that use late binding of tasks to pilots instead of to resources.

Experiment 3 (Table~\ref{table:exp_swift}, third row) uses an execution
strategy with one pilot for each resource. The size of each pilot is the
total number of tasks that need to be executed divided by the total number of
pilots that have been scheduled. The duration of a pilot is the time required
to execute the complete BoT on the number of cores of that pilot. This
strategy minimizes the number of pilots and cores required for each resource,
maintaining maximal concurrency only in the best-case scenario in which all
the pilots become active at the same time. In the worse-case scenario, the
BoT is executed on a single Pilot with as much concurrency as allowed by the
number of cores of that pilot.

The execution strategy of Experiment 3 is consistent with the insight gained
from Experiment 1 and 2 and with previous experimental
results~\cite{turilli2016integrating}. As each pilot can execute all the
tasks of the given BoT, \(TTC\) should not depend on the last pilot becoming
active, as in Experiments 1 and 2, but on the first one (and on how many
pilots become active after the first pilot).

\begin{figure}
  \centering
  \includegraphics[width=\linewidth]{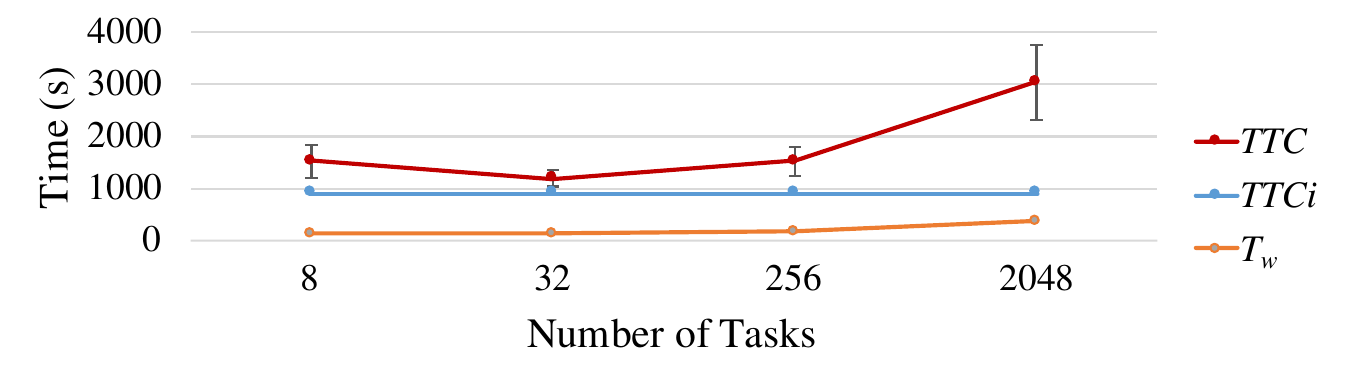}
  \caption{\B{Experiment 3}. Average \(TTC\) and \(T_w\) for bags of
           8, 32, 256, 2,048 tasks executed via AIMES middleware on Stampede
           and Gordon, improved for all the BoT when compared to those of
           Experiments 1 and 2. \(TTC_i\) is 1,200~s for all the BoTs. Note:
           The Y-axis range is between 0 and 6,000 s in this figure, between
           0 and 120,000 s in Fig.~\ref{fig:exp_swift_1}
           and~\ref{fig:exp_swift_2}.}\label{fig:exp_aimes_3}
\end{figure}

Fig.~\ref{fig:exp_aimes_3} confirms the reduction of average \(TTC\) for
Experiment 3 and, as expected, a corresponding reduction in \(T_w\). The
third row of Table~\ref{table:es_efficiency} shows an increased efficiency of
the execution strategy used for Experiment 3 when compared to those used for
Experiment 1 and 2.

Barring dedicated or largely underutilized queues, queue time of multi-tenant
production resources  is mostly unpredictable; it depends on per user
policies and on the state of the queue at every point in
time~\cite{wolski2003experiences}. Queue waiting time tends to vary over time
for each user, and varies differently across resources. The execution
strategy of Experiment 3 responds to these fluctuations by submitting pilots
to two resources and late binding tasks only to active pilots.

The difference between Experiment 4 and Experiment 3's execution strategy is
that the former uses four XSEDE resources.  As the number of resources used
increases, the time for the first pilot to become active should decrease.
Intuitively, binding tasks only to active pilots should improve with
increasing number of resources. However, this improvement must be traded off
against the increased overhead of binding tasks to a larger number of
resources.

\begin{figure}
  \centering
  \includegraphics[width=\linewidth]{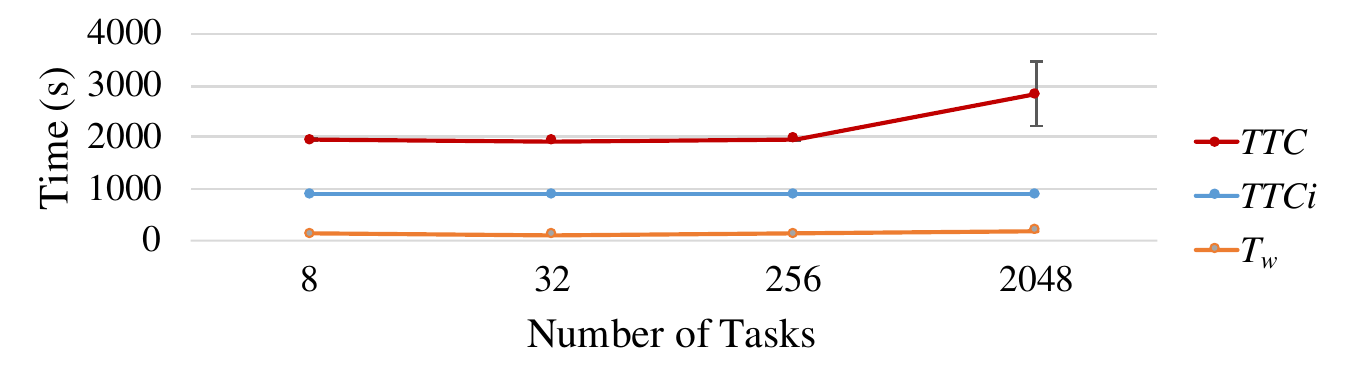}
  \caption{ \B{Experiment 4}. Average \(TTC\) and \(T_w\) for bags of 8, 32,
            256, 2,048 tasks executed via AIMES middleware on Stampede,
            Gordon, SuperMIC, and Comet. \(TTC_i\) is 1,200~s for all the
            BoTs. Compared to Experiment 3, average \(TTC\) improves for bags
            of 2,048 tasks but worsens for all the other BoT\@; average
            \(T_w\) is instead analogous across the BoT
            sizes.\label{fig:exp_aimes_4}}
\end{figure}

The fourth row of Table~\ref{table:es_efficiency} shows an improvement in the
average \(P_{ES}\) for BoTs with 2048 tasks but a worsening for BoTs with
less tasks than 2048. Our analysis shows that this is due to increased
overheads of the AIMES middleware in scheduling across four resources, and
the very short queuing time of the pilots of both Experiment 3.

Fig.~\ref{fig:exp_aimes_4} shows average \(T_w\), analogous to those in
Fig.~\ref{fig:exp_aimes_3}, indicating that the differences in average
\(TTC\) between Experiments 4 and 3 depend on \(T_x\). The communication
between pilots and the AIMES Unit Manager via a database service
(Fig.~\ref{subfig:aimes-arch}) adds overheads to \(T_x\): the more resources
are used, the higher is the time taken to communicate during the execution of
each task~\cite{merzky2015radical}.

Our experiments were executed in four intervals across three months to sample
the behavior of the resources' queues in different periods.
Fig.~\ref{fig:exp_aimes_3} and~\ref{fig:exp_aimes_4} show that we experienced
very short \(T_w\) compared to the historical annual average queue time
recorded by XDMoD~\cite{xdmod_url}. In presence of longer \(T_w\), Experiment
3's execution strategy could perform worse than Experiment 4's due to the
reduced interplay between the queue times of just 2 resources, which would be
consistent with previous data~\cite{turilli2016integrating}.

% ---------------------------------------------------------------------------
\subsection{Integrated: AIMES and Swift}\label{ssec:exp_integration}

We used the insight gained from the comparison of alternative execution
strategies to perform experiments with the integrated Swift and AIMES
middleware.

\begin{figure}
  \centering
  \includegraphics[width=\linewidth]{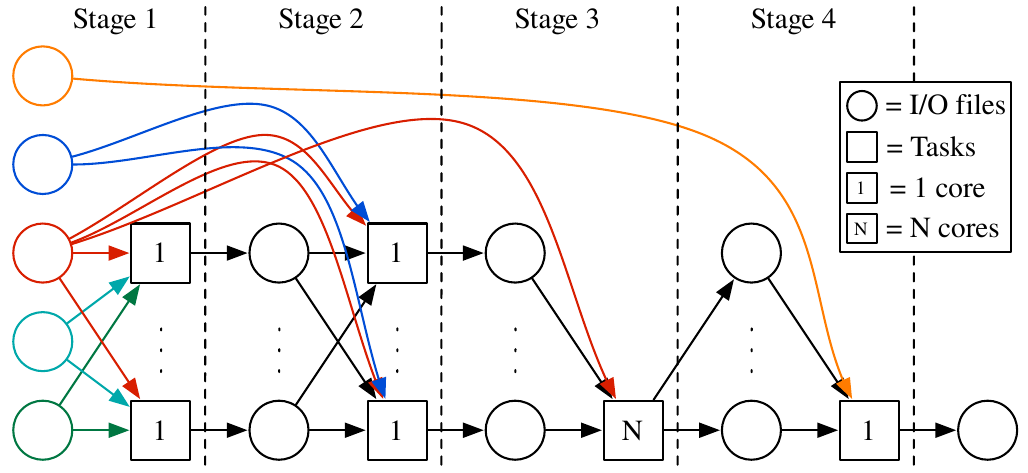}
  \caption{Molecular Dynamics workflow~\cite{balasubramanian2016ensemble}. 4
    stages process five input files returning an output file, Stages 1 and 2
    perform concurrent simulations and Stages 3 and 4 perform aggregated
    analysis.}\label{fig:workflow}
\end{figure}

Fig.~\ref{fig:workflow} shows a workflow developed within the
ExTASY~\cite{balasubramanian2016extasy} project and used to execute a
Simulation Analysis Loop pattern. The workflow is used to model a solvated
alanine dipeptide molecule containing 2,881 atoms. Each simulation executes
the Amber MD Engine for 0.6 ps followed by analysis of all
simulations~\cite{balasubramanian2016ensemble}.  The workflow comprises four
stages, two for the simulations and two for the analyses. Stage 1 has
\(N\) 1-core simulation tasks, each taking the three input files of the
workflow and returning one output file. Stage 2 is comprised of \(N\) 1-core
simulation tasks, each taking one output files of Stage 1 and two input files
of the workflow, and returning one output file. Stage 3 consists of one MPI
analysis task with N cores, taking all Stage 2 output files and one input
file of the workflow, and returns N output files. Stage 4 executes one
analysis task with one core, taking all output files of Stage 3 and one input
file of the workflow, and returns a single output file.

We used Synapse~\cite{merzky2016synapse} to generate emulated tasks that have
the performance of ExTASY
tasks.  We then executed the ExTASY workflow with 256 to 2,048 simulations,
using Experiment 4's execution strategy. We used up to five resources: four
from XSEDE and Blue Waters. We measured average \(TTC\) at each scale and
utilized the analytical tools we developed for both AIMES and Swift to
measure the \(TTC\) time components.

\begin{figure}
  \centering
  \includegraphics[width=\linewidth]{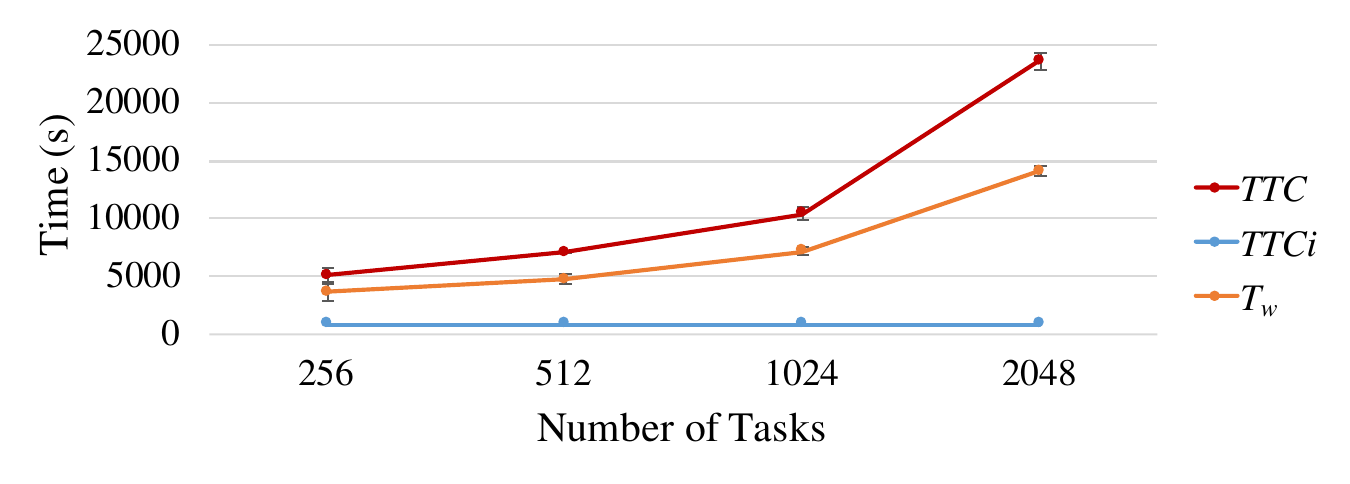}
  \caption{ Average \(TTC\) and \(T_w\) for workflows
            (Fig.~\ref{fig:workflow}) with 256, 1,024, and 2,048 simulations,
            executed via the AIMES and Swift integrated system on Stampede,
            Gordon, SuperMIC, Comet, and Blue Waters. The aggregated
            \(TTC_i\) for all the stages is an average of 870~s as profiled
            by Synapse. \(T_w\) grows proportional to pilot size, \(TTC\)
            shows \(T_x\) overheads proportional to the number of the
            simulations.}\label{fig:workflow_results}
\end{figure}

Fig.~\ref{fig:workflow_results} shows a progressive increase of both \(TTC\)
and \(T_w\) averages. Our analysis shows that \(TTC\) increases mainly due to
the time taken to stage input and output files between the user's workstation
and the resource of the pilot on which each task is executed. While input and
output files are on average few hundred KB, their number increases with the
scale of the workflow. The progressive increase of \(T_w\) is due to a
corresponding increase in the pilots' average queue time. This likely depends
on the size of the requested pilots: the larger the number of tasks of the
workflow, the larger the size of each pilot on each resource
(Table~\ref{table:exp_swift}, execution strategy of Experiment 4).

Fig.~\ref{fig:workflow_results} also shows small error bars for both \(TTC\)
and \(T_w\). For \(TTC\), this indicates that network latency, staging
mechanisms, and the concurrency of task execution are consistent across
executions, and that the execution strategy of Experiment 4 keeps both the
mean and variance of queue waiting time low. Similar to Experiment 3 and 4,
this is obtained by binding to active pilots and a reduction in the typical
time for the activation of the first pilot.

Rows 4 and 5 of Table~\ref{table:es_efficiency} show that the same execution
strategy can perform differently depending on the type of workload executed.
The same strategy performs well with the BoT used for Experiment 4 but poorly
with the workflow used for the integrated experiments. This depends on file
staging, required by the integrated experiments but not by Experiment 4: data
are staged in and out of remote resources for each task during execution.
This adds overhead to \(TTC\) but it is necessary as, currently, the AIMES
task scheduler does not schedule tasks to a pilot based on whether their
input data are already available to that pilot.

The explicit definition of the execution strategy allows us to isolate the
decisions that affect the performance of the execution for the \(TTC\)
metric. For example, an execution strategy with a single resource and an
early binding of tasks to that resource could be used for the integrated
experiments. This would avoid intermediate data staging between the user's
workstation and the resource, and enable a comparison between the two
alternative strategies.

% ---------------------------------------------------------------------------
% CONCLUSIONS
% ---------------------------------------------------------------------------
\section{Conclusions}\label{sec:discussions}

Distributed execution of heterogeneous workloads on heterogeneous resources
opens a large problem space with both conceptual and implementation
challenges. We focused on the issues of deciding among alternative ways to
distribute executions and on comparing their performance using the notion of
an execution strategy. We described, compared, and integrated AIMES and
Swift; then we used these systems to analyze the performance of four
execution strategies with two types of workloads, at different scales, and on
multiple and diverse resources.

Our comparison of AIMES and Swift uncovered their architectural and
functional analogies and differences (\S\ref{sec:archs} and
\S\ref{sec:comparison}). The same core features enable the two systems to
perform distributed executions but because of their diverse capabilities,
they enact these executions differently. Task-based workload description and
pilot-based abstraction of resource capabilities enable distributed
execution, but different ways to bind tasks to resources or schedule tasks to
pilots leads to different realizations of the execution.  {\it In this way,
we reiterate that the challenge is not engineering distributed execution but
characterizing and measuring alternative ways to distribute those
executions.}

Our contribution is to advance and extend the concept of execution
strategy~\cite{turilli2016integrating}.  We integrated Swift and AIMES into a
system specifically designed to execute, characterize, and measure
alternative execution strategies (\S\ref{sec:integration}). From an
implementation perspective, our integrated prototype confirmed the benefits
of a RESTful API but, more importantly, it showed how to reconcile diverse
approaches to workload management and disjoint state models, without
re-engineering AIMES or Swift.

Our experiments compared and measured the performance of four alternative
execution strategies (Table~\ref{table:exp_swift}). We contributed a
definition of the performance of execution strategies (\(P_{ES}\)) based on
observed and ideal \(TTC\). The performance differences between Experiments
1--2 and 3--4 show the relevance of resource availability on distributed
executions (Table~\ref{table:es_efficiency}).  We explained this by showing
the dominance of \(T_w\) and importance of responding to different queue time
across multiple resources (\S\ref{ssec:exp_aimes_swift}).

The performance differences between Experiment 4 and the experiments using
the integrated AIMES and Swift system showed how the performance of execution
strategies also depends on the characteristics of the executed workload
(\S\ref{ssec:exp_integration}). By introducing data staging, the execution
strategy of Experiment 4 performed more poorly that the integrated
experiments. We clarified that this was not due to the size of the file
transferred, but a byproduct of how late binding to pilots is implemented.
This was reflected in the values of \(P_{ES}\) for the integrated experiments
(Table~\ref{table:es_efficiency}).

This work indicates several directions for future research. Conceptually, the
notion of execution strategy needs to be generalized to provide greater
quantitative insight, and its use extended to more systems and more use
cases.  To test of the validity of the execution strategy abstraction and to
stress its generality, we plan to explore its integration with PANDA-WMS, the
primary workload management system for the ATLAS project. Investigating these
and other tradeoffs will be the subject of future research.

The AIMES and Swift integration will be further developed.  Currently, the
integrated prototype enables execution but offers limited introspective
capabilities. Swift's and AIMES' state models cannot be fully coordinated due
to the limitations in how information is shared between the two systems.
Thus, no error management, fault-tolerance, or shared logging is available
during execution. Extending the RESTful API to exchange information while
maintaining the separation of concerns between the two systems will be a core
requirement for developing a production-grade integration between AIMES and
Swift.

The integration of Swift and AIMES provides a preliminary case study of
effective integration between independent tools that enable distributed
computing.  We believe more case studies are needed to understand how to
reduce the number of redundant and competing tools and thus create a
sustainable software ecosystem. \\

% ---------------------------------------------------------------------------
% CONTRIBUTIONS AND ACKNOWLEDGMENTS
% ---------------------------------------------------------------------------

{\footnotesize {\bf Contributions \& Acknowledgments:} MT and SJ conceived
this paper. MT led design of the experiments, primarily wrote the paper (with
AM, MW, DSK, and SJ contributing), and performed the analysis. YNB and AM
developed the Swift- AIMES provider; AM developed the RESTful API and
integrated experiments. YNB contributed to experiments 1 and 2, and MTH
performed experiments 3 and 4.  MT, DSK and SJ edited the paper. This work is
funded by DOE ASCR DE-FG02-12ER26115, DE-SC0008617, DE-SC0008651 and NSF
ACI-1253644. We acknowledge access to XSEDE computational facilities (TG-
MCB090174) and Blue Waters (NSF 1516469). We thank Mihael Hategan for his
support of the work in this paper. Some work by Katz was supported by NSF
while at the NSF; any opinion, finding, and conclusions or recommendations
expressed in this material are those of the author(s) and do not necessarily
reflect the views of the NSF.\par}

% ---------------------------------------------------------------------------
% REFERENCES
% ---------------------------------------------------------------------------
\vspace*{1em}
\newcommand{\BIBdecl}{\setlength{\itemsep}{0.25 em}}
\bibliographystyle{IEEEtran}
\bibliography{aimes}

\end{document}